\documentclass[twocolumn,showpacs,preprintnumbers,amsmath,amssymb]{revtex4}
\usepackage[dvips]{graphicx}
\usepackage{color}

\def\cor#1{{#1}}
\def\ket#1{  \left\vert  #1   \right\rangle   }
\def\bra#1{  \left\langle  #1   \right\vert   }

\begin{document}

\title{A Quantum Model for Autonomous Learning Automata}

\author{Michael Siomau}
\email{m.siomau@gmail.com}

\affiliation{Physics Department, Jazan University, P.O.~Box 114,
45142 Jazan, Kingdom of Saudi Arabia}

\date{\today}

\begin{abstract}
The idea of information encoding on quantum bearers and its
quantum-mechanical processing has revolutionized our world and
brought mankind on the verge of enigmatic era of quantum
technologies. Inspired by this idea, in present paper we search for
advantages of quantum information processing in the field of machine
learning. Exploiting only basic properties of the Hilbert space,
superposition principle of quantum mechanics and quantum
measurements, we construct a quantum analog for Rosenblatt's
perceptron, which is the simplest learning machine. We demonstrate
that the quantum perceptron \cor{is superior} its classical
counterpart in learning capabilities. In particular, we show that
the quantum perceptron is able to learn an arbitrary (Boolean)
logical function, perform the classification on previously unseen
classes and even recognize the superpositions of learned classes --
the task of high importance in applied medical engineering.
\end{abstract}


\maketitle

\section{\label{sec:1} Introduction}

During last few decades, we have been witnessing unification of
quantum physics and classical information science that resulted in
constitution of new disciplines -- quantum information and quantum
computation \cite{Nielsen:00,Georgescu:13}. While processing of
information, which is encoded in systems exhibiting quantum
properties suggests, for example, unconditionally secure quantum
communication \cite{Gisin:02} and superdense coding
\cite{Vedral:02}, computers that operate according to the laws of
quantum mechanics offer efficient solving of problems that are
intractable on conventional computers \cite{Childs:10}. Having
paramount practical importance, these announced technological
benefits have indicated the main directions of the research in the
field of quantum information and quantum computation, somehow
leaving aside other potential applications of quantum physics in
information science. So far, for instance, very little attention has
been paid on possible advantages of quantum information processing
in such areas of modern information science as machine learning
\cite{Kecman:01} and artificial intelligence \cite{Russel:09}.
\cor{Using standard quantum computation formalism, it has been shown
that machine learning governed by quantum mechanics has certain
advantages over classical learning \cite{Menneer:95, Andrecut:02,
Kouda:05, Zhou:06, Zhou:07, Manzano:09}. These advantages, however,
are strongly coupled with more sophisticated optimization procedure
than in the classical case, and thus require an efficiently working
quantum computer \cite{Ladd:10} to handle the optimization. This
paper, in contrast, presents a new approach to machine learning,
which, in the simplest case, does not require any optimization at
all.}

Our focus is on perceptron, which is the simplest learning machine.
Perceptron is a model of neuron that was originally introduced by
Rosenblatt \cite{Rosenblat:57} to perform visual perception tasks,
which, in mathematical terms, result in solution of the
\textit{linear classification problem}. There are two essential
stages of the perceptron functioning: supervised learning session
and new data classification. During the first stage, the perceptron
is given a labeled set of examples. Its task is of inferring weights
of a linear function according to some error-correcting rule.
Subsequently, this function is utilized for classification of new
previously unseen data.

In spite of its very simple internal structure and learning rule,
the perceptron's capabilities are seriously limited
\cite{Minsky:69}. Perceptron can not provide the classification, if
there is an overlap in the data or if the data can not be linearly
separated. It is also incapable of learning complex logical
functions, such as XOR function. Moreover, by its design, the
perceptron can distinguish only between previously seen classes and,
therefore, can not resolve the situation when the input belongs to
none of the learned classes, or represents a superposition of seen
classes.

In this paper we show that all the mentioned problems can be, in
principle, overcome by \cor{a quantum analog for perceptron}. There
are also two operational stages for the quantum perceptron. During
the learning stage all the data are formally represented through
quantum states of physical systems. This representation allows
expanding the data space to a physical Hilbert space. It is
important to note, that there is no need to involve real physical
systems during this stage. Thus, the learning is essentially a
classical procedure. The subject of the learning is a set of
positive operator valued measurements (POVM) \cite{Nielsen:00}. The
set is constructed by making superpositions of the training data in
a way that each operator is responsible for detection of one
particular class. This procedure is linear and does not require
solving equations or optimizing parameters. When the learning is
over, there are two possibilities to achieve the required
classification of new data. First, new data are encoded into the
states of real quantum systems, which are measured by detectors
adjusted in accordance with the learned POVM. Second, new data may
be formally encoded into the states of quantum systems and processed
with the POVM. Both mentioned ways allow to achieve the
classification.

This paper is organized as follows. In the next section, we first
overview the classical perceptron and discuss the origin of the
restrictions on its learning capabilities. After this, in
Section~\ref{sec:2b}, we introduce the quantum perceptron and show
its properties. We demonstrate, in Section~\ref{sec:3}, three
examples of how the quantum perceptron \cor{is superior} its
classical counterpart in learning capabilities: complex logical
function learning, classification of new data on previously unseen
classes and recognition of superpositions of classes. We conclude in
Section~\ref{sec:4}.

\section{\label{sec:2} Basic Constructions}

\subsection{\label{sec:2a} Rosenblatt's Perceptron}

Operational structure of the classical perceptron is simple. Given
an input vector $\textbf{x}$ (which is usually called a feature
vector) consisting of $n$ features, perceptron computes a weighted
sum of its components $f(\textbf{x}) = \sum_i a_i x_i$, where
weights $a_i$ have been previously learned. The output from a
perceptron is given by $o = {\rm sign} (f(\textbf{x}))$, where ${\rm
sign(...)}$ is the Heaviside function
\begin{equation}
 \label{sign}
 {\rm sign} (y) = \{
 \begin{array}{cc}
  +1 & y>0 \\
  -1 & y \leq 0 \\
 \end{array} \, .
\end{equation}
Depending on the binary output signal $o \in \{ +1,-1\}$, the input
feature vector $\textbf{x}$ is classified between two feature
classes, one of which is associated with output $o=+1$ and the other
with output $o=-1$.

As we have mentioned above, the perceptron needs to be trained
before its autonomous operation. During the training, a set of P
training data pairs $\{ \textbf{x}_i, d_i, i=1,...,P \}$ is given,
where $\textbf{x}_i$ are the $n$-dimensional feature vectors and
$d_i$ are desired binary outputs. Typically, at the beginning of the
learning procedure the initial weights $a_i$ of the linear function
are generated randomly. When a data pair is chosen from the training
set, the output $o_i = {\rm sign} (f(\textbf{x}_i))$ is computed
from the input feature vector $\textbf{x}_i$ and is compared to the
desired output $d_i$. If the actual and the desired outputs match
$o_i = d_i$, the weights $a_i$ are left without change and the next
pair from the data set is taken for the analysis. If $o_i \ne d_i$,
the weights $a_i$ of the linear function are to be changed according
to the error-correcting rule $\textbf{a}^\prime = \textbf{a} +
\epsilon \textbf{a} = \textbf{a} + (d_i - o_i) \textbf{x}_i$, which
is applied hereafter and until the condition $o_i = d_i$ is met.

\begin{figure}
\begin{center}
\includegraphics[scale=0.57]{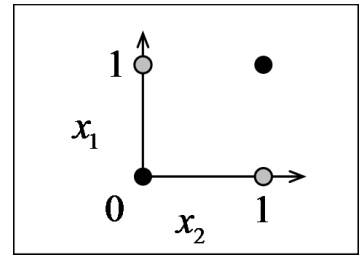}
\caption{The feature space of XOR function is two-dimensional and
discrete (each feature takes only values 0 and 1). There is no line
(a one-dimensional hyperplane) that separates black and grey points.
Classical perceptron is incapable of classifying the input feature
vectors and, therefore, can not learn XOR function.}
 \label{fig-2}
\end{center}
\end{figure}

The training procedure has clear geometric interpretation. The
weights $a_i$ of the linear function define a $\left( n-1
\right)$-dimensional hyperplane in the $n$-dimensional feature
space. The training procedure results in a hyperplane that divides
the feature space on two subspaces, so that each feature class
occupies one of the subspaces. Due to this interpretation, the
origin of the restrictions on learning capabilities of the classical
perceptron becomes visible: a hyperplane that separates the two
classes may not exist. The simplest example of two classes that can
not be linearly separated is XOR logical function of two variables,
which is given by the truth table
\begin{equation}
 \label{XOR}
  \begin{array}{ccccc}
  x_1 & \; 0 & \; 0 & \; 1 & \; 1 \\
  x_2 & \; 0 & \; 1 & \; 0 & \; 1 \\
  f   & \; 0 & \; 1 & \; 1 & \; 0 \\
  o   & -1 & +1 & +1 & -1 \\
 \end{array} \, .
\end{equation}
A schematic representation of this function in the two-dimensional
feature space is shown in Fig.~\ref{fig-2}.

There are, however, limitations on the learning capabilities of the
perceptron even in the case when the separating hyperplane exists.
As we mentioned above the hyperplane divides the feature space on
two subspaces, in spite of the fact that the feature classes occupy
two particular hypervolumes. This enforces the classification on the
two learned classes even so the given feature is essentially
different from the classes, i.e. form a new class.

It is very important to note that certain tasks undoable by
Rosenblatt's perceptron, such as complex logical functions learning
and classifying data with an overlap, can be performed in the
framework of more sophisticated classical learning models, for
example, by support vector machines \cite{Kecman:01}. However, these
classical implementations always demand \textit{nonlinear}
optimization, which complicates rapidly with growth of the feature
space. This effect is known as the curse of dimensionality of the
classical learning models \cite{Kecman:01}. In the next section, we
present a new model for the learning machine, which, however, is
\textit{linear}, but \cor{is superior} Rosenblatt's perceptron in
its learning capabilities.

\subsection{\label{sec:2b} \cor{A quantum analog for Rosenblatt's perceptron}}

As its classical counterpart, quantum perceptron is to be trained to
perform the classification task. Suppose, we are given a set of $K$
training data pairs consisting of feature vectors  $\{\textbf{x}_k,
d_k, k=1,...,K \}$ with the desired binary outputs $d \in \{ +1,-1
\}$; and each feature vector consists of $n$ features $\textbf{x} =
\{ x_1, x_2, ..., x_n \}$. \cor{Let us suppose that each feature is
restricted in a certain interval, so that all features can be
normalized to the unit interval $x_k^\prime \in \left[ 0,1\right]$
for $k=1,...,n$. This allows us to represent the input feature
vectors through the states of a (discrete) $2^n$-dimensional quantum
system, so that $\ket{\textbf{x}} =
\ket{x_1^\prime,x_2^\prime,...,x_n^\prime}$. With this quantum
representation we have extended the classical $n$-dimensional
feature space to $2^n$-dimensional Hilbert space of the quantum
system. We shall drop "primes" hereafter assuming that the features
are normalized.}

\cor{Let us construct a projection operator
$\ket{\textbf{x}}\bra{\textbf{x}}$ for each given feature vector
$\ket{\textbf{x}}$. With the help of these projectors, let us define
two operators
\begin{eqnarray}
 \label{operators-def}
 \nonumber
P_{-1} & = & \frac{1}{N_{-1}} \sum_{d=-1}
\ket{\textbf{x}}\bra{\textbf{x}} \, ,
\\[0.1cm]
P_{+1} & = & \frac{1}{N_{+1}} \sum_{d=+1}
\ket{\textbf{x}}\bra{\textbf{x}} \, ,
\end{eqnarray}
where $N_{-1}$ and $N_{-1}$ are normalization factors. All feature
vectors that correspond to the output $d=-1$ are summed in the
operator $P_{-1}$, while all feature vectors corresponding $d=+1$
are collected in $P_{+1}$.} The construction of these operators
concludes the learning procedure.

There are only four possibilities of how the operators $P_{-1}$ and
$P_{+1}$ may be related:

\textit{A.} Operators $P_{-1}$ and $P_{+1}$ are orthogonal $P_{-1}
P_{+1} =0$ and form a complete set $P_{-1} + P_{+1} = I$, where $I$
is the identity operator. This means that there was no overlap
between the training data, and the two classes $P_{-1}$ and $P_{+1}$
occupy the whole feature space. As the result any input feature
vector can be classified between the two classes with no mistake.
This situation can be simulated in principle by the classical
perceptron.

\textit{B.} Operators $P_{-1}$ and $P_{+1}$ are orthogonal $P_{-1}
P_{+1} =0$, but do not form a complete set $P_{-1} + P_{+1} \ne I$.
This is an extremely interesting case. The third operator must be
defined as $P_{0} = I - P_{-1} - P_{+1}$ to fulfill the POVM
competence requirement. The operator $P_{0}$ is, moreover,
orthogonal to $P_{-1}$ and $P_{+1}$, because $P_{-1} P_{+1} =0$.
When operating autonomously, the quantum perceptron generates three
outputs $d \in \{ +1, 0, -1 \}$, namely that the feature vector
belongs to the one of the previously seen classes $d \in \{ +1,-1
\}$ or it is essentially different from the learned classes $d = 0$
-- it belongs to a new previously unseen class. The classification
on previously unseen classes is an extremely hard learning problem,
which can not be done by classical perceptron neither by the most of
the classical perceptron networks \cite{Kecman:01}. Quantum
perceptron is capable of performing this task. Moreover, there will
be no mistake in the classification between the three classes
because of the orthogonality of the operators $P_{-1}, P_{+1}$ and
$P_0$.

\textit{C.} Operators $P_{-1}$ and $P_{+1}$ are not orthogonal
$P_{-1} P_{+1} \ne 0$, but form a complete set $P_{-1} + P_{+1} =
I$. In this case all the input data can be classified between the
two classes with some nonzero probability of mistake. This is the
case of probabilistic classification, which can not be done by the
classical perceptron, although can be performed by more
sophisticated classical learning models.

\textit{D.} The most general case is when operators $P_{-1}$ and
$P_{+1}$ are not orthogonal $P_{-1} P_{+1} \ne 0$ and do not form a
complete set $P_{-1} + P_{+1} \ne I$. One again defines the third
operator $P_0 = I - P_{-1} - P_{+1}$, which this time is not
orthogonal to $P_{-1}$ and $P_{+1}$. In this situation, quantum
perceptron classifies all the input feature vectors on three
classes, one of which is a new class, with some nonzero probability
of mistake. This situation can not be simulated by the classical
perceptron.

The quantum perceptron learning rule may have the following
geometric interpretation. In contrast to the classical perceptron,
which constructs a hyperplane separating the feature space on two
subspaces, quantum perceptron constructs two (hyper-)volumes in the
physical Hilbert space. These volumes are defined by the POVM
operators (\ref{operators-def}). During the autonomous functioning,
the POVM operators project the given feature vector $\ket{\psi}$ to
one of the volumes (or to the space unoccupied by them) allowing us
to perform the desired classification. For example, if $\bra{\psi}
P_{-1} \ket{\psi} \neq 0$, while $\bra{\psi} P_{+1} \ket{\psi} = 0$
and $\bra{\psi} P_{0} \ket{\psi} = 0$, the feature vector
$\ket{\psi}$ belongs to the class $d= -1$, and the probability of
misclassification equals zero. If, in contrast, $\bra{\psi} P_{-1}
\ket{\psi} \neq 0$, $\bra{\psi} P_{+1} \ket{\psi} \neq 0$ and
$\bra{\psi} P_{0} \ket{\psi} = 0$, the feature vector belongs to the
two classes with degrees defined by the corresponding expectation
values $\bra{\psi} P_{-1} \ket{\psi}$ and $\bra{\psi} P_{+1}
\ket{\psi}$. In the latter situation, one may perform a
probabilistic classification according to the expectation values.

\cor{We would like to stress that the construction of the operators
(\ref{operators-def}) is no way unique. There may be more
sophisticated ways to construct the POVM set in order to ensure a
better performance of the learning model for a classification
problem at hand. In fact, our construction is the simplest liner
model for a quantum learning machine. Only in this sense the
presented quantum perceptron is the analog for Rosenblatt's
perceptron, while their learning rules are essentially different. }

As we mentioned in the Introduction, there are two ways to achieve
the desired classification with the POVM. One may get real physical
systems involved or use the POVM operators as purely mathematical
instrument. In order of clarity, the advantages of the first of
these approaches will be discusses in Section~\ref{sec:3a} on
particular examples, while in the rest of the next section we use
the quantum perceptron as pure mathematical tool.

\section{\label{sec:3} Applications}

In spite of the extreme simplicity of its learning rule, quantum
perceptron may perform a number of tasks infeasible for classical
(Rosenblatt) perceptron. In this section we give three examples of
such tasks. We start with logical function learning. Historically,
the fact that classical perceptron can not learn an arbitrary
logical function was the main limitation on the learning
capabilities of this linear model \cite{Minsky:69}. We show that
quantum perceptron, in contrast, is able of learning an arbitrary
logical function irrespective of its kind and order. In
Section~\ref{sec:3b}, we show that quantum perceptron can, in
certain cases, perform the classification without previous training,
the so-called unsupervised learning task. Classical perceptron, in
contrast, can not perform this task by construction. Finally, in
Section~\ref{sec:3c} we show that quantum perceptron may recognize
superpositions of previously learned classes. This task is of
particular interests in applied medical engineering, where
simultaneous and proportional myoelectric control of artificial limb
is a long desired goal \cite{Jiang:12}.

\subsection{\label{sec:3a} Logical Function Learning}

Let us consider a particular example of logical function -- XOR,
which is given by the truth table (\ref{XOR}). During the learning
session, we are given a set of four training data pairs
$\{\textbf{x}_i, d_i, i=1,...,4 \}$, where the feature vector
consists of two features $\textbf{x} \in \{ x_1, x_2 \}$, and the
desired output $d \in \{ +1,-1 \}$ is a binary function. Let us
represent the input features through the states of a two-dimensional
quantum system -- qubit, so that each feature is given by one of the
basis states $\ket{x_i} \in \{ \ket{0}, \ket{1} \}$ for $i=1,2$,
where $\{ \ket{0}, \ket{1} \}$ denotes the computational basis
\cor{for each feature}. In the above representation, the feature
vector $\textbf{x}$ is given by one of the four two-qubit states
$\ket{x_1, x_2}$. Following the procedure, which is described in
Section~\ref{sec:2b}, the POVM operators are constructed as
\begin{eqnarray}
 \label{pure-operators}
 \nonumber
P_{-1} & = & \ket{0,0} \bra{0,0} + \ket{1,1} \bra{1,1} \, ,
\\[0.1cm]
P_{+1} & = & \ket{0,1} \bra{0,1} + \ket{1,0} \bra{1,0} \, .
\end{eqnarray}

During its autonomous operation, quantum perceptron may be given
four basis states $\ket{x_1, x_2} \in \{ \ket{0,0}, \ket{0,1},
\ket{1,0}, \ket{1,1} \}$ as inputs. Since $\bra{x_1, x_2} P_{-1}
\ket{x_1, x_2} \neq 0$ only for $\ket{x_1, x_2} \in \{ \ket{0,0},
\ket{1,1} \}$, these states are classified to $d= -1$, while the
other two states $\{ \ket{0,1}, \ket{1,0}\}$ are classified to $d=
+1$. The fact that the operators $P_{-1}$ and $P_{+1}$ are
orthogonal ensures zero probability of misclassification, while the
completeness of the set of operators guarantees classification of
any input. Conclusively, the quantum perceptron has learned XOR
function.

The successful XOR function learning by quantum perceptron is the
consequence of the representation of the classical feature vector
$\textbf{x}$ through the two-qubit states. In the classical
representation, the feature vectors can not be linearly separated on
a plane, see Fig.~\ref{fig-2}. In the quantum representation, four
mutually orthogonal states $\ket{x_1, x_2} $ in the four-dimensional
Hilbert space can be separated on two classes in an arbitrary
fashion. This implies that an arbitrary logical function of two
variables can be learned by quantum perceptron. For example,
learning of logical AND function leads to the construction of
operators $P_{-1} = \ket{0,0}\bra{0,0} + \ket{0,1}\bra{0,1} +
\ket{1,0}\bra{1,0}$ and $P_{+1} = \ket{1,1} \bra{1,1}$. Moreover, an
arbitrary logical function of an arbitrary number of inputs
(arbitrary order) also can be learned by quantum perceptron, because
the number of inputs of such a function growth exponentially as
$2^n$ with the order of the function $n$ and exactly as fast as
dimensionality of the Hilbert space that is needed to represent the
logical function.

\cor{In the above discussion the need to use real quantum systems
has not emerged.} Let us now consider a situation, when one can
benefit from utilizing real quantum systems. Let us slightly modify
the problem of XOR learning. In real-life learning tasks the
training data may be corrupted by noise \cite{Kecman:01}. In some
cases, noise may lead to overlapping of the training data, which
result in misclassification of feature vectors during the training
stage and during further autonomous functioning. For example, if,
during the XOR learning, there is a finite small probability
$\delta$ that feature $x_1$ takes a wrong binary value, but the
other feature and the desired output are not affected by noise,
after a big number of trainings (which are usually required in case
of learning from noisy data),  the POVM operators are given by
\begin{eqnarray}
 \label{noisy-operators}
{P'}_{-1} & = & (1-\delta) \left( \ket{0,0} \bra{0,0} + \ket{1,1}
\bra{1,1} \right)
\nonumber \\[0.1cm]
& & \hspace{0.5cm} + \; \delta \left(\ket{0,1} \bra{0,1} + \ket{1,0}
\bra{1,0} \right) \, ,
\nonumber \\[0.1cm]
{P'}_{+1} & = & (1-\delta) \left( \ket{0,1} \bra{0,1} + \ket{1,0}
\bra{1,0}\right)
\nonumber \\[0.1cm]
& & \hspace{0.5cm} + \; \delta \left( \ket{0,0} \bra{0,0} +
\ket{1,1} \bra{1,1}\right) \, .
\end{eqnarray}

Operators ${P'}_{-1}$ and ${P'}_{+1}$ are not orthogonal ${P'}_{-1}
{P'}_{+1} \ne 0$ in contrast to operators (\ref{pure-operators}),
\cor{but still form a complete set.}  This means that during the
autonomous operation of the quantum perceptron, the input feature
vectors can be misclassified. Nevertheless, each feature is
classified between the two classes and, on average, most of the
feature vectors are classified correctly. This means that quantum
perceptron simulates XOR function with a degree of accuracy given by
$1 - \delta$.

If we use real physical systems to encode feature vectors during
autonomous functioning of the perceptron and measure the states of
the systems with experimental setup adjusted in accordance with the
POVM (\ref{noisy-operators}), we can perform a probabilistic
classification. Moreover, we can exactly (in probabilistic sense)
reproduce fluctuations that have been observed during the training.
In certain sense such learning is too accurate and may be of use in
some cases. Anyway, classical perceptron can not do any similar
task.

\cor{It is, however, important to note that practical simulation of
quantum perceptron with real physical systems may not be always
possible. In Section \ref{sec:2b} we discussed situations when
operators $P_{-1}$ and $P_{+1}$ do not form a compete set, and
constructed the third operator $P_0 = I - P_{-1} - P_{+1}$. It is
possible in principle that the constructed operator $P_0$ is
negative, i.e. unphysical. This means that the classification
problem at hand can not be physically simulated with our linear
model, although the problem may be treated mathematically with the
quantum perceptron approach.}

\cor{In this section we have seen how quantum representation and
quantum measurements contribute to advanced learning abilities of
the quantum perceptron. Even without these features, however,
quantum perceptron is superior its classical counterpart in learning
capabilities due to specific algebraic structure of the POVM
operators. In the following sections we provide two examples, where
advanced learning relays only on the structure of the POVM set.}

\subsection{\label{sec:3b} Unsupervised Learning}

The (supervised) learning stage, has been embedded into quantum
perceptron by analogy with classical perceptron. Surprisingly,
however, that the learning rule of the quantum perceptron allows to
perform learning tasks beyond supervised learning paradigm. Suppose,
for example, that we are given an unlabeled set of feature vectors
and need to find a possible structure of this set, i.e. we need to
answer whether there are any feature classes in the set. The
following protocol allows us to resolve such an unsupervised
learning task under certain conditions.

\cor{Being given the first feature vector $\ket{\textbf{x}_1}$ from
the set, let us define two classes with the POVM operators
\begin{eqnarray}
 \label{operators-unsup-learn}
 \nonumber
P^{(0)}_{-1} & = & \ket{\textbf{x}_1}\bra{\textbf{x}_1} \, ,
\\[0.1cm]
P^{(0)}_{+1} & = & I - P^{(0)}_{-1} \, .
\end{eqnarray}
where $I$ is the identity operator. Here, the class $d=+1$ is
formally defined as "not $d=-1$". The next given feature vector
$\ket{\textbf{x}_2}$ is tested to belong to one of these classes. If
$\bra{\textbf{x}_2} P^{(0)}_{-1} \ket{\textbf{x}_2} >
\bra{\textbf{x}_2} P^{(0)}_{+1} \ket{\textbf{x}_2}$, the feature
vector $\ket{\textbf{x}_2}$ is close enough to $\ket{\textbf{x}_1}$
and thus belongs to class $d=-1$. In this case the POVM operators
(\ref{operators-unsup-learn}) are updated to
\begin{eqnarray}
 \label{operators-unsup-learn-first-it-2}
 \nonumber
P^{(1)}_{-1} & = &  \ket{\textbf{x}_1}\bra{\textbf{x}_1} +
\ket{\textbf{x}_2}\bra{\textbf{x}_2}\, ,
\\[0.1cm]
P^{(1)}_{+1} & = & I - P_{-1} \, .
\end{eqnarray}
If, in contrast, $\bra{\textbf{x}_2} P^{(0)}_{-1} \ket{\textbf{x}_2}
\geq \bra{\textbf{x}_2} P^{(0)}_{+1} \ket{\textbf{x}_2}$, the
feature vector $\ket{\textbf{x}_2}$ is distant sufficiently from
$\ket{\textbf{x}_1}$ and therefore can be assigned a new class
$d=+1$. Due to the first representative of the $d=+1$ class, we may
update the formal definition of the $P^{(0)}_{+1}$ introducing a new
POVM set
\begin{eqnarray}
 \label{operators-unsup-learn-first-it-2}
 \nonumber
P^{(1)}_{-1} & = & \ket{\textbf{x}_1}\bra{\textbf{x}_1} \, ,
\\[0.1cm]
P^{(1)}_{+1} & = & \ket{\textbf{x}_2}\bra{\textbf{x}_2} \, .
\end{eqnarray}
This procedure is repeated iteratively until all the feature vectors
are classified between the two classes $d=-1$ and $d=+1$.}

The above protocol will work if only there are at least two feature
vectors $\ket{\textbf{x}}$ and $\ket{\textbf{y}}$ in the given
feature set such as $\bra{\textbf{x}} (I-2P) \ket{\textbf{x}} \geq
0$, where $P = \ket{\textbf{y}} \bra{\textbf{y}}$. In the opposite
case, unsupervised learning within the protocol is not possible.
Moreover, the classification crucially depends on order of examples,
because first seen feature vectors define the classes. This
situation is, however, typical for unsupervised learning models
\cite{Kecman:01}. \cor{To reduce the dependence of the
classification on the order of the feature vectors appearance, it is
possible to repeat the learning many times taking different order of
the input feature vectors, and compare the results of the
classification.} In spite of the above limitations, the unsupervised
classification can be in principle performed by the quantum
perceptron, while this task is undoable for the classical
perceptron.

\subsection{\label{sec:3c} Simultaneous and Proportional Myoelectric Control}

The problem of signal classification has found remarkable
applications in medical engineering. It is known that muscle
contraction in human body is governed by electrical neural signals.
These signals can be acquired by different means \cite{Parker:04},
but are typically summarized into so-called electromyogram (EMG). In
principle, processing the EMG, one may predict muscular response to
the neural signals and subsequent respond of the body. This idea is
widely used in many applications, including myoelectric-controlled
artificial limb, where the surface EMG is recorded from the remnant
muscles of the stump and used, after processing, for activating
certain prosthetic functions of the artificial limb, such as hand
open/close \cite{Jiang:12}.

Despite decades of research and development, however, none of the
commercial prostheses is using pattern classification based
controller \cite{Jiang:12}. The main limitation on successful
practical application of pattern classification for myoelectric
control is that it leads to very unnatural control scheme. While
natural movements are continuous and require activations of several
degrees of freedom (DOF) simultaneously and proportionally,
classical schemes for pattern recognition allow only sequential
control, i.e. activation of only one class that corresponds to a
particular action in one decision \cite{Jiang:12}. Simultaneous
activation of two DOFs is thus recognized as a new class of action,
but not as a combination of known actions. Moreover, all these
classes as well as their superpositions must be previously learned.
This leads to higher rehabilitation cost and more frustration of the
user, who must spend hours in a lab to learn the artificial limb
control.

Recently, we have taken quantum perceptron approach to the problem
of simultaneous and proportional myoelectric control
\cite{Siomau:13}. We considered a very simple control scheme, where
two quantum perceptrons were controlling two degrees of freedom of
the wrist prosthesis. We took EMG signals with corresponding angles
of the wrist position from an able-bodied subject who performs wrist
contractions. For the training we used only those EMG that
correspond to the activation of a single DOF. During the test, the
control scheme was given EMG activating multiple DOFs. We found that
in 45 of 55 data blocks of  the actions were recognized correctly
with accuracy exceeding $73\%$, which is comparable to the accuracy
of the classical schemes for classification.

In the above example, \cor{we used a specific representation of the
feature vectors. Since the features (i.e. the neural signals) are
real and positive numbers there was no need to expand the feature
space. Moreover, in general it is not possible to scale a given
feature on the unit interval, because the neural signals observed
during the learning and autonomous functioning may differ
significantly in amplitude, and \textit{a priori} scaling may lead
to misapplication of the artificial limb. Therefore, the amplitude
of a signal was normalized over amplitudes from all the channels to
ensure proportional control of the prosthesis. In fact, the specific
structure of the POVM set was the only feature of the quantum
perceptron that we used. With this feature alone we were able to
recognize 4 original classes observed during the training and 4 new
(previously unseen) classes that correspond to simultaneous
activation of two DOF.} In general, within the above control scheme,
$n$ quantum perceptrons are able to recognize $2n$ original classes
with $(2n)!/[2(2n-2)!]-n$ additional two-class superpositions of
these classes. In contrast, $n$ classical perceptrons may recognize
only $2n$ classes, which were seen during the learning. \cor{The
advantage of the quantum perceptron over the classical perceptron
can be understood from the geometric interpretation discussed in
Section~\ref{sec:2b}. While $n$ classical perceptrons construct $n$
hyperplanes in the feature space, which separate the feature space
on $2n$ non-overlapping classes, $n$ quantum perceptrons build $n$
hypervolumes, which may not fill the whole feature space and may
overlap.}

\section{\label{sec:4} Conclusion}

Bridging between quantum information science and machine learning
theory, we showed that the capabilities of an autonomous learning
automata can be dramatically increased using the quantum information
formalism. We have constructed the simplest linear quantum model for
learning machine, which, however, \cor{is superior} its classical
counterpart in learning capabilities. \cor{Due to the quantum
representation of the feature vectors, the probabilistic nature of
quantum measurements and the specific structure of the POVM set, the
quantum perceptron is capable of learning an arbitrary logical
function, performing probabilistic classification, recognizing
superpositions of previously seen classes and even classifying on
previously unseen classes. Since all classical learning models track
back to Rosenblatt's perceptron, we hope that the linear quantum
perceptron will serve as a basis for future development of
practically powerful quantum learning models, and especially in the
domain of nonlinear classification problems.}

\begin{acknowledgments}
This project is supported by KACST.
\end{acknowledgments}

\end{document}